\documentclass[aps, prd, amsmath, floats, floatfix,  superscriptaddress, nofootinbib]{revtex4}

\usepackage{graphicx,color}
\usepackage{latexsym}
\usepackage{amsmath,amssymb}        
\usepackage[draft=false]{hyperref}
\usepackage{mathrsfs}
\usepackage{comment}
\usepackage{soul}
\definecolor{orange}{rgb}{1,0.5,0}
\DeclareSymbolFontAlphabet{\mathrsfs}{rsfs}
\DeclareMathAlphabet{\mathcal}{OMS}{cmsy}{m}{n}

\usepackage{ulem}

\newcommand{\beq}  {\begin{equation}}
\newcommand{\eeq}  {\end{equation}}

\newcommand{\beqa}    {\begin{eqnarray}}
\newcommand{\eeqa}    {\end{eqnarray}}

\newcommand{\no}    {\\ \nonumber}

\def\physrep{{\em Phys. Rep.}  }

\def\apj{{\em Astrophys. J.  }}
\def\apjl{{\em Astrophys. J. Lett. }}
\def\mnras{{ Mon. Not. Roy. Astron. Soc.  }}
\def\jcap{{JCAP }}

\begin{document}
\title{Quantum corrections as a Bound for Detecting Self-Interacting Ultralight Dark Matter}

\author{Jae-Weon Lee}
\email{scikid@jwu.ac.kr}
\affiliation{Department of Electrical and Electronic Engineering, Jungwon University, 85 Munmu-ro, Goesan-eup, Goesan-gun, Chungcheongbuk-do, 28024, Korea.}

\author{Chueng-Ryong Ji}\email{ji@ncsu.edu}
\affiliation{Department of Physics and Astronomy, North Carolina State University, Raleigh, North Carolina 27695-8202}

\begin{abstract}
We investigate the implications of interactions between ultralight dark matter (ULDM) and Standard Model fermions for the radiative stability of the ULDM potential. By matching the heavy fermion onto a low-energy scalar effective field theory, we derive the one-loop threshold corrections to the ULDM mass and self-interaction couplings. We interpret these corrections in terms of radiative naturalness conditions. For the parameter region motivated by cosmological and galactic considerations, the quadratic correction provides the strongest naturalness criterion, while the direct and cubic-induced quartic corrections remain subdominant. Our results illustrate how radiative effects can constrain 
the parameter space satisfying our 
radiative-naturalness criterion based on threshold corrections.
\end{abstract}

\maketitle

\section{Introduction}
The ultralight dark matter (ULDM) model is a well-motivated alternative to cold dark matter (CDM), in which dark matter consists of bosons with an extremely small mass, typically $m \sim 10^{-22} {\rm eV}$, forming a Bose–Einstein condensate on galactic scales (for reviews, see Refs. \cite{2009JKPS...54.2622L,2014ASSP...38..107S,2014MPLA...2930002R,Chavanis_2011,2016PhR...643....1M,Hui:2016ltb}). ULDM can be treated as
a classical field
$\phi$.
This framework is also referred to as fuzzy dark matter, BEC dark matter, scalar field dark matter, ultralight axion, or wave-$\psi$ dark matter \cite{Baldeschi:1983mq,Sin:1992bg,Lee:1995af,1993ApJ...416L..71W,2000PhRvL..84.3037N,Arbey:2001qi,Hu:2000ke,Peebles:2000yy,2009PhLB..671..174M,2000PhRvD..62j3517S,Alcubierre:2001ea,2012PhRvD..86h3535P,2009PhRvL.103k1301S,Fuchs:2004xe,Matos:2001ps,2001PhRvD..63l5016N,Boehmer:2007um,Eby:2015hsq}.

A key property of ULDM is its large de Broglie wavelength, $\lambda_{dB}=\hbar/(m v)\sim\rm{kpc}$, which sets the characteristic size of the smallest galaxies and helps alleviate CDM small-scale problems such as the core–cusp, missing-satellite, and satellite-plane issues \cite{Lee:2008ux,2009Natur.460..717V,2022JCAP...12..033P,Salucci:2002nc,1996ApJ...462..563N,2003MNRAS.340..657D,2003IJMPD..12.1157T}. ULDM has also been applied to black-hole-related puzzles, including the $M$–$\sigma$ relation and the final parsec problem \cite{Lee:2015yws,koo2024final,Bromley:2023yfi}. In the non-interacting (fuzzy) limit, quantum pressure arising from the uncertainty principle counteracts gravity~\cite{Koo:2025jkx}.

However, fuzzy dark matter is strongly constrained by Lyman-$\alpha$ forest observations, which typically require $m \gtrsim 10^{-21}\mathrm{\rm eV}$ \cite{Irsic:2017yje,Armengaud:2017nkf,Koo:2025jkx}. 
 Zimmermann et al.~\cite{Zimmermann:2024xvd} obtained
$
m > 2.2\times10^{-21}\ {\rm eV}
\qquad (95\%~{\rm C.L.})
$
from the stellar kinematics of Leo II, while Teodori et al.~\cite{Teodori:2025rul} found  tension with dwarf-galaxy observations for
$
m \lesssim 5\times10^{-21}\ {\rm eV}.
$
Introducing self-interactions into the ULDM framework \cite{Lee:1995af,Boehmer:2007um,Chavanis_2011} has been proposed as a way to relax these bounds and enlarge the viable parameter space \cite{Dave:2023wjq,Hartman:2021upg,Shapiro:2021hjp}. More recently, self-interacting ULDM (SIULDM) has also been discussed in connection with the Hubble tension and with broader particle-physics puzzles such as neutrino masses and the electroweak scale \cite{Lee:2025roi,Lee:2024rdc}.

Another advantage of
SIULDM is that
it may be
probed experimentally through its interactions
 with
ordinary matter.
When the effective mass of $\phi$
induced by the self-interaction potential $U(\phi)$
exceeds the Hubble parameter $H$, the background field $\phi$ begins to oscillate. Numerous efforts have been made to detect these oscillations using precision probes such as atomic clocks~\cite{Arvanitaki:2014faa} and gravitational-wave detectors~\cite{Chowdhury:2023xvy}.
Even if ULDM has no tree-level self-interaction, it can acquire an effective self-interaction through one-loop corrections induced by
the interactions with
ordinary matter.
Radiative corrections to ultralight scalar potentials induced by couplings
to Standard Model or other fermionic fields have been well studied in several
contexts
~\cite{Arvanitaki:2014faa,Arvanitaki_2010,Aghaie:2026bqq,Murgui:2023kig}.

In the present work, we focus on the cosmologically motivated regime of repulsive SIULDM with $\tilde{m}=m/\lambda^{1/4}$
of order a  few ${\rm eV}$ and formulate the heavy-fermion effect through threshold matching onto a low-energy scalar 
effective field theory (EFT), showing that the scalar-mass naturalness condition 
~\cite{Delaunay:2025pho}
automatically suppresses both the direct and cubic-induced quartic corrections relative to the tree-level quartic coupling.

In this paper, 
we summarize some of the cosmological and galactic constraints on SIULDM, use them to define the benchmark parameter region, and then derive radiative naturalness criteria for its couplings to ordinary matter.
In Section II, we briefly review cosmological and galactic constraints on the SIULDM particle mass $m$ and the quartic self-coupling $\lambda$. In Section III, we derive the one-loop corrections to 
$m$ and $\lambda$ induced by interactions with ordinary matter and
 evaluate their radiative naturalness in the cosmologically motivated SIULDM parameter region.

\section{COSMOLOGICAL AND GALACTIC CONSTRAINTS ON SIULDM}

In this section,
we briefly review the cosmological 
and galactic constraints on SIULDM.
Cosmological and galactic observations typically 
favor the range
 $1~{\rm eV} \lesssim\tilde{m} \lesssim 10~{\rm eV}$ ~\cite{Li:2013nal,Hartman:2021upg},
where $\tilde{m}\equiv m/\lambda^{1/4}$
is the typical energy scale of
$\phi$.

In this work the SIULDM field is represented as a real scalar field $\phi$ with the action:  
\beq
S
=
\int d^4x\sqrt{-g}
\left[
-\frac{R}{16\pi G}
-\frac{1}{2}g^{\mu\nu}
\partial_\mu\phi
\partial_\nu\phi
-U(\phi)
\right],
\eeq
where $U(\phi)$
is the potential of $\phi$. 
We concentrate on the quartic self-interaction with $\lambda>0$, which is the highest even-power term that remains renormalizable. In this study, we do not consider a cosine-type potential, as it effectively yields an attractive quartic interaction.

Therefore, in this work $U(\phi)$ is defined as  
\beq
U(\phi) = \frac{m^2}{2} \phi^2 + \frac{\lambda }{4} \phi^4.
\label{U1}
\eeq  
Then, the evolution of the scalar field is governed by
\beq
\Box\phi
-
\frac{dU}{d\phi}
=
0,
\eeq
where $\square$ is the covariant d'Alembertian operator with gravity.  
We can decompose
the field $\phi$ into 
a homogeneous
background part
and a perturbation.
For a spatially homogeneous background field $\phi(t)$, the equation of motion reduces to
\beq
\ddot{\phi} + 3H\dot{\phi} + m^2\phi + \lambda \phi^3 = 0,
\label{eom}
\eeq
which governs the cosmological evolution of the background scalar field, where the dot denotes a derivative with respect to cosmic time.
(We will use
natural units $\hbar=1=c$, unless otherwise stated.)
To derive the relevant redshifts, we will use the relation between the redshift $z$ and the temperature of the universe $T$,
$
1+z(T)=\frac{T}{T_0},$
where $T_0=2.35\times10^{-4} {\rm eV}$ is the present temperature of the universe.

For a coherently oscillating scalar, we define the effective equation of state parameter as
$w \equiv \langle p^\phi \rangle / \langle \rho^\phi \rangle$,
where $\langle p^\phi \rangle$ is the time-averaged pressure of $\phi$.
For the oscillating real scalar field with a potential of the form $U(\phi)\propto \phi^n$, the time-averaged energy density $\rho^\phi$ behaves as a perfect fluid characterized by an effective equation of state $w = \frac{n-2}{n+2}$ \cite{PhysRevD.28.1243}. 
Accordingly, the evolution of the background field can be divided into three distinct phases, depending on which term in $U(\phi)$ dominates.

In Phase I ($z \ge z_{osc}$), Hubble friction keeps the field nearly frozen, so $\phi$ effectively acts as early dark energy with $w \simeq -1$. In our scenario, however, its energy density  remains subdominant during this stage. As the universe expands further, the field starts to roll down the potential and undergoes coherent oscillations for $z<z_{osc}$.

In Phase II ($z_m \le z < z_{osc}$), the quartic self-interaction term in $U(\phi)$ dominates over the quadratic mass term. During this epoch, the oscillating field redshifts like radiation, with an effective equation of state $w \simeq 1/3$ and an energy density scaling as $\rho^\phi \sim T^4$.

In Phase III ($z < z_m$), the quadratic mass term overtakes the quartic interaction, and the oscillating scalar field behaves as pressureless matter, corresponding to $w = 0$, and
$\rho^\phi \sim T^3$.

The transition from Phase II to Phase III occurs at the temperature $T_m$, defined by the equality between the quartic and quadratic terms in the potential,
$
m^2\phi^2/2\simeq \lambda \phi^4/4,
$
which gives
\beq
\phi=\phi_m
\simeq \frac{\sqrt{2}m}{\sqrt{\lambda}}\simeq \sqrt{2}\frac{\tilde{m}^2}{m}.
\label{phim}
\eeq
If we require $\phi_m$ to be smaller than the Planck mass, we get 
$m\gtrsim 5.9\times 10^{-26}{\rm eV}$ 
for $\tilde{m}\simeq 10~{\rm eV}$.

At matter-radiation equality in Phase III, the energy density of the field $\rho^\phi$ is taken to be
\beq
\label{rhophieqT}
\rho^\phi_{eq}\simeq \rho_{\text{rad}}(T_{eq}) = \frac{\pi^2}{30} g_* T_{eq}^4,
\eeq
where $g_* = 3.36$ is the number of relativistic degrees of freedom. 
From this we can
obtain $\rho^\phi$
at other epochs.
To treat the oscillation of $\phi$ as CDM, 
the quartic term in $U(\phi)$ must be smaller than the quadratic term, at least at the epoch of matter-radiation equality at $z_{eq}$. Equivalently, the field amplitude at that time, $\phi=\phi_{eq}$, must satisfy $\phi_{eq}<\phi_m= \sqrt{2}\frac{\tilde{m}^2}{m}$. One can then infer the present-day ULDM density parameter.
\beq
\Omega_\phi \simeq \frac{{\frac{m^2\phi_{eq}^2}{2}}\left(\frac{T_{\text{0}}}{T_{\text{eq}}}\right)^3}{3H_0^2 m^2_P}
\lesssim \frac{ \tilde{m}^4}{3  H_0^2 m_P^2} \left(\frac{T_{{0}}}{T_{\text{eq}}}\right)^3
\eeq
which leads to~\cite{Boudon:2022dxi} 
$\tilde{m}  \gtrsim 0.77~{\rm eV}$,
where $H_0$ is the Hubble constant and $\Omega_\phi\simeq 0.26.$

In Phase III, the field behaves as CDM, and the energy density scales as $\rho^\phi\propto T^3$. Considering
both  the quadratic and the quartic contributions,
the energy density
at $T=T_m$ is therefore
\beq
\rho^\phi_{m}
\simeq 2~m^2 \phi_m^2/2 = 2\frac{m^4}{\lambda}=2\tilde{m}^4
=\rho^\phi_{eq} \left( \frac{T_m}{T_{eq}}\right)^3,
\eeq
from which we obtain
\beq
\label{Tm}
T_m=\left(\frac{2\tilde{m}^4}{\rho^\phi_{eq}}\right)^{1/3} T_{eq}.
\eeq
The SIULDM field safely behaves as CDM provided that
$
T_m>T_{eq}\simeq 0.79~{\rm eV},
$
which corresponds to $\tilde{m}> { 0.68}~{\rm eV}$~\cite{Li:2013nal,Boudon:2022dxi}. 
For $\tilde{m}=O(1)~ {\rm eV}$ one can see that $T_m=O(1)~{\rm eV}$.

In Phase II, the equation of motion
(Eq. (\ref{eom})) is well approximated by
\beq
\ddot{\phi}+\lambda \phi^3=0,
\eeq
and the characteristic oscillation timescale is of $O(1/(\sqrt{\lambda}\phi))$.
The field begins to oscillate at $\phi=\phi_{osc}$ and $T=T_{osc}$, when the force induced by the potential overcomes the Hubble friction. This condition is
approximately expressed as
$
3H\dot{\phi}_{osc}\simeq
H\sqrt{\lambda} \phi^2_{osc}
\simeq\lambda \phi^3_{osc}$.
Therefore, at this moment,
$
H\simeq \sqrt{\lambda}{\phi_{osc}}
\simeq
{T_{osc}^2}/{m_P},
$
which leads to
\beq
\label{Tosc}
T_{osc}\simeq \lambda^{1/4}\sqrt{{m_P}{\phi_{osc}}}
\simeq \left( \frac{ \rho^\phi_{osc} m_P^2}{\phi_{osc}^2}\right)^{1/4},
\eeq
where 
$
\rho^\phi_{osc}
\equiv\rho^\phi(T_{osc})\simeq \lambda \phi^4_{osc},
$
and $m_P=2.4\times 10^{18}~{\rm GeV}$ is the reduced Planck mass.
Note that this relation already carries an order-one uncertainty. In deriving the analytic estimates below, we also neglect
the residual variation of both $g_*$ and $g_{*s}$. Therefore, constraints involving $\phi_{osc}$ should be regarded as rough estimates.

Next, we relate the energy density at matter–radiation equality to $\phi_{osc}$. 
Using the scaling $\rho^\phi\propto T^4$ in Phase II and $\rho^\phi\propto T^3$ in Phase III, we find
\beq
\rho^\phi_{eq}\simeq\rho^\phi_{osc} \left(\frac{T_m}{T_{osc}}\right)^4 \left(\frac{T_{eq}}{T_{m}}\right)^3
=\frac{\rho^\phi_{osc} T_m T_{eq}^3}{T_{osc}^4}
=\frac{\phi^2_{osc} T_m T_{eq}^3}{m_P^2},
\label{rhophieq}
\eeq
which yields
\beq
\label{phiosc1}
\phi_{osc}\simeq \sqrt{\frac{\rho^\phi_{eq} m_P^2}{T_mT^3_{eq}}}
=\frac{ {\pi}^{4/3} {m_P} \left(\frac{{g_*} {T_{eq}} }{{\tilde{m}}}\right)^{2/3}}{2^{1/6} 30^{2/3}},
\eeq
where we
used 
Eq. (\ref{Tm}) and Eq. (\ref{rhophieqT}).

The analytic cosmological requirements above give \(\tilde m\gtrsim0.77\,{\rm eV}\), consistent with the observationally motivated scale of order \(1\) {\rm eV} or larger.
These numerical values should be regarded as approximate benchmark
estimates rather than universal sharp limits, since the variation of
the effective relativistic and entropy degrees of freedom has been
neglected in the analytic treatment.

Now, we consider constraints from galactic dynamics,
where we focus on the perturbations of the
field about the background field. 
In the nonrelativistic regime relevant to galaxies, the spacetime-dependent scalar field $\phi(t,\mathbf{x}) $ can be expressed as:  
\beq
\phi(t, \mathbf{x}) = \frac{1}{\sqrt{2m}} \left[e^{-i m t} \psi(t, \mathbf{x}) + e^{i m t} \psi^*(t, \mathbf{x})\right].
\eeq   

In natural units, time-averaging 
of $\phi^2$ gives $\langle\phi^2\rangle=|\psi|^2/m$, while the mass density is $\rho=m|\psi|^2$. Restoring $\hbar$ and $c$,
  the macroscopic wave function $\psi$ then satisfies the nonlinear Schr\"odinger-Poisson equations (SPE):  
\beq
i\hbar\partial_t\psi
=
-\frac{\hbar^2}{2m}\nabla^2\psi
+
mV\psi
+
\frac{3\lambda\hbar^3}{4m^2c}
|\psi|^2\psi, \quad \nabla^2 V = 4\pi G \rho,
\eeq  
where the DM mass density is $\rho = m |\psi|^2$, and $V$ is the gravitational potential.  
For a real scalar field, time-averaging of the quartic interaction produces the factor $3/4$ appearing in the nonlinear Schr\"odinger interaction.

In the Thomas–Fermi (TF) limit, where the kinetic term can be ignored,  the  stationary ground state solution is given by
\beq
|\psi(r)|^2=\frac{|\psi(0)|^2 R_{TF}}{\pi r} sin\left(\frac{\pi r}{R_{TF}}\right),
\eeq
where
the Thomas–Fermi radius is 
\beq
R_{\rm TF}
=
\sqrt{
\frac{3\pi\hbar^3\lambda}
{16cGm^4}
},
\eeq
which can be interpreted as the minimum core size of small galaxies
and as half
of the Jeans length~\cite{1985MNRAS.215..575K,Chavanis_2011,LeeJi2025QuantumScales}
\beq
\lambda_J
=
2R_{\rm TF}
=
\sqrt{
\frac{3\pi\hbar^3\lambda}
{4cGm^4}
}= 1.197~{\rm kpc}\left(\frac{10~{\rm eV}}{\tilde{m}}\right)^{2}.\eeq

\begin{figure*}[tb]
\centering
\includegraphics[width=0.49\textwidth]{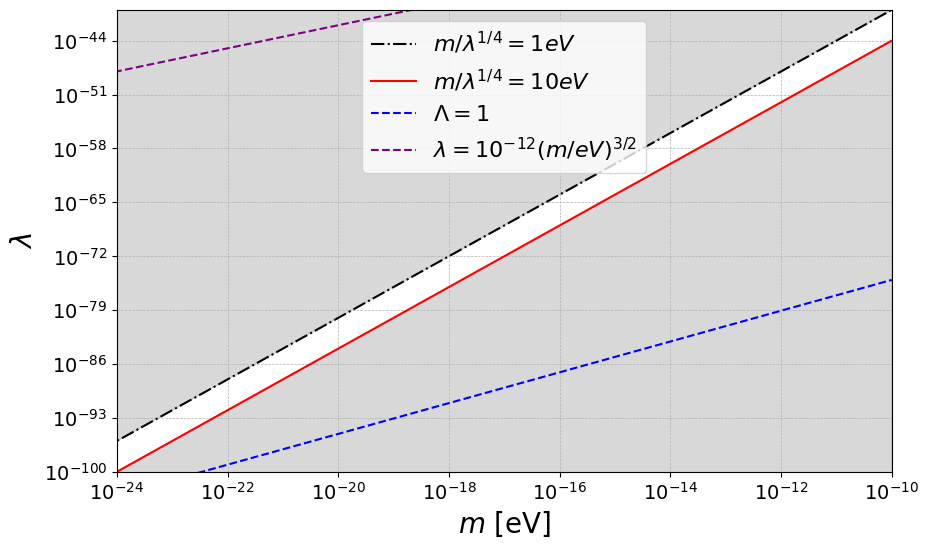}
 \caption{Selected cosmological and galactic constraints in the parameter plane $(m, \lambda)$, where the 
 shaded regions are disfavored.
}
 \label{lambda}
\end{figure*}

Dynamical analyses of classical dwarf spheroidal galaxies generally favor dark-matter cores
on scales ranging from several hundred parsecs to a few
kiloparsecs
~\cite{Amorisco_2012,Diez_Tejedor_2014}. 
Motivated by these results, we identify the observed core scale with
$R_{\rm TF}$ up to an order-unity factor and adopt the representative range
\beq
0.5~{\rm kpc} \lesssim R_{\rm TF} \lesssim 5~{\rm kpc},
\eeq
which corresponds to
\beq
3.46~{\rm eV} \lesssim \tilde m \lesssim 10.94~{\rm eV}.
\eeq

The TF regime for a repulsively self-interacting boson star is characterized by the dimensionless coupling ~\cite{Lee:1995af}
\beq
\Lambda \equiv \frac{\lambda M_{\rm Pl}^{2}}{4\pi m^{2}} \gg 1.
\eeq
Taking the boundary of the TF regime to be defined by $\Lambda=1$, one obtains
\beq
\lambda_{\rm TF}(m)=\frac{4\pi m^{2}}{M_{\rm Pl}^{2}}
\simeq 8.44\times10^{-56}
\left(\frac{m}
{\rm eV}\right)^{2},
\eeq
where $
M_{\rm Pl}$ is the
unreduced Planck mass
$
M_{\rm Pl}
=
1.22\times10^{28}~{\rm eV}
$.

Observations of merging galaxy clusters, such as the Bullet Cluster, place an upper limit on the self-interaction cross section per unit mass, $\sigma / m_{\mathrm{DM}} \lesssim 1\rm{cm}^2/\rm{g}$ \cite{Tulin:2017ara}. Translating this bound to the quartic self-interaction of ULDM yields roughly the constraint \cite{Garcia:2023abs}
$
\lambda \lesssim 10^{-12}\left(\frac{m}{1\rm{eV}}\right)^{3/2}$.

Considering these results, we adopt the broad benchmark range \(1\,{\rm eV}\lesssim\tilde m\lesssim10\,{\rm eV}\) and use \(\tilde m=5\,{\rm eV}\) as a representative value in the following analysis.

\section{Effective potential and One-loop corrections}

In this section, we investigate the radiative effects induced by Yukawa interactions between the ULDM scalar field and ordinary matter~\cite{Arvanitaki:2014faa}. Since the fermion mass $m_f$ is much larger than the characteristic energy scale of the ULDM sector, the heavy fermion should not be treated as an active degree of freedom at arbitrarily low energies. Instead, we match the full theory containing both the scalar field and the fermion onto a low-energy scalar effective field theory (EFT) near the fermion threshold $\mu \simeq m_f$. The effects of the heavy fermion are then encoded in the coefficients of  scalar operators. We use these threshold corrections to quantify the radiative naturalness of the ULDM mass and self-interaction.

We consider a real scalar field $\phi$ coupled to a massive Standard Model fermion $\Psi$ through a Yukawa interaction. The renormalized Lagrangian is written as
\beq
\mathcal{L}
=-\frac{1}{2}(\partial_\mu\phi)(\partial^\mu\phi)
-
U_R(\phi)
+
\bar{\Psi}(i\not{\partial}-m_f)\Psi
-
y\phi\bar{\Psi}\Psi,
\eeq
where $y$ is a dimensionless positive Yukawa coupling. 

The Yukawa interaction explicitly breaks the discrete symmetry $\phi\rightarrow-\phi$. Consequently, even if the scalar potential contains only even powers of $\phi$ at tree level, radiative corrections generally generate both odd and even powers of the scalar field. We therefore write the renormalized scalar potential in the general form
\beq
U_R(\phi)
=
t_R\phi
+
\frac{m_R^2}{2}\phi^2
+
\frac{a_R}{3}\phi^3
+
\frac{\lambda_R}{4}\phi^4,
\eeq
where $t_R$ is the renormalized tadpole coefficient, $m_R$ is the renormalized scalar mass parameter, $a_R$ is the renormalized cubic coupling, and $\lambda_R$ is the renormalized quartic coupling.
The corresponding counterterm potential is
\beq
U_{\rm CT}(\phi)
=
\delta t_{\rm CT}\phi
+
\frac{\delta m^2_{\rm CT}}{2}\phi^2
+
\frac{\delta a_{\rm CT}}{3}\phi^3
+
\frac{\delta\lambda_{\rm CT}}{4}\phi^4.
\eeq
The counterterms cancel the ultraviolet divergences generated by loop corrections. Their finite parts also specify the renormalization prescription used to define the renormalized parameters.

At the heavy-fermion threshold, the finite one-loop contributions to the scalar one-, two-, three-, and four-point functions are absorbed into the coefficients of the corresponding operators in the low-energy scalar EFT. 
We will follow the one-loop matching procedure discussed by Braathen et al.~\cite{Braathen:2018htl}.
The full one-loop effective potential
for $\phi$ is written as
\beq
U_{\rm full}(\phi)
=
U_R(\phi)
+
U_{\rm CT}(\phi)
+
U^{(1)}(\phi)
\eeq
where $U^{(1)}(\phi)$ denotes the one-loop contribution.
In this work we will focus on the
fermionic contribution to $U^{(1)}(\phi)$.

For a constant scalar background $\phi$, the Yukawa interaction changes the fermion mass to
\beq
M_f(\phi)
=
m_f + y\phi.
\eeq

Using dimensional regularization in the $\overline{\rm MS}$ scheme, 
for the color-singlet fermion the fermionic Coleman-Weinberg contribution to the one-loop effective potential is
~\cite{Sher:1988mj,Aghaie:2026bqq}
\beq
U_f^{(1)}(\phi)
=
-\frac{M_f^4(\phi)}{16\pi^2}
\left[
\log\left(
\frac{M_f^2(\phi)}{\mu^2}
\right) -
\frac{3}{2}
\right],
\eeq
where $\mu$ is the renormalization scale.
The overall minus sign is the standard fermionic contribution
arising from the Grassmann functional determinant.

The expansion of this expression in powers of $\phi$ is valid when the Yukawa-induced change in the fermion mass is small compared with $m_f$. We therefore require
\beq
|y\phi|
\ll
m_f.
\eeq

For convenience, we define $L\equiv
\log\left(
\frac{m_f^2}{\mu^2}
\right).$
Expanding the fermionic one-loop potential through fourth order in $\phi$ gives
\beqa
U_f^{(1)}(\phi)
\simeq
&&
\frac{ym_f^3}{4\pi^2}
\left(1-L\right)\phi
+
\frac{y^2m_f^2}{8\pi^2}
\left(
1-3L
\right)\phi^2
\no
&&
-\frac{y^3m_f}{4\pi^2}
\left(L+\frac{2}{3}\right)\phi^3
-\frac{y^4}{16\pi^2}
\left(L+\frac{8}{3}
\right)\phi^4,
\eeqa
where the term proportional to $\phi$ is a loop-induced tadpole. The term proportional to $\phi^2$ gives a radiative correction to the scalar mass parameter. The term proportional to $\phi^3$ generates a cubic scalar interaction even if the renormalized cubic interaction is chosen to vanish at a reference scale. Finally, the term proportional to $\phi^4$ gives a direct fermion-loop correction to the quartic self-interaction.

This work focuses on the electron because it permits a direct comparison with existing precision constraints on the electron-mass coupling $d_{m_e}$. The extension to neutrino couplings is model dependent
~\cite{Murgui:2023kig}
because the neutrino mass operator depends on whether neutrino masses are Dirac or Majorana and on the underlying mass-generation mechanism. We therefore restrict the numerical benchmark to the electron and leave neutrino couplings for a separate study.
For the electron,
$
m_f= m_e
= 0.511~{\rm MeV}
$
whereas the characteristic SIULDM energy scale motivated by the cosmological analysis in the previous section is approximately
$\tilde m
\simeq
5~{\rm eV}$.
Therefore,
$
\frac{m_f}{\tilde m}
\simeq
10^5
$
and evaluating the full-theory effective potential directly at $\mu=\tilde m$ would give
\beq
L=
\log\left(
\frac{m_f^2}{\tilde m^2}
\right)
\simeq
23.1.
\eeq
The appearance of this relatively large logarithm reflects the presence of two widely separated energy scales in a single fixed-order effective potential.
To avoid carrying this large logarithm in the threshold calculation, we perform the matching near the heavy-fermion scale and choose
$\mu=m_f$.
At the matching scale,
$
L=
\log\left(
\frac{m_f^2}{\mu^2}
\right)
=0.
$
The finite fermionic one-loop contribution at $\mu=m_f$ therefore becomes
\beq
U_f^{(1)}(\phi)\big|_{\mu=m_f}
=
\frac{ym_f^3}{4\pi^2}\phi
+
\frac{y^2m_f^2}{8\pi^2}\phi^2 
-\frac{y^3m_f}{6\pi^2}\phi^3
-\frac{y^4}{6\pi^2}\phi^4.
\eeq
For energies above the matching scale, the scalar and fermion are both retained explicitly in the full theory. For processes characterized by external momenta $p_i\ll m_f$, the heavy fermion can instead be integrated out and its virtual effects represented by local scalar operators.

Formally, integrating out the fermion means performing the functional integral over the fermion fields. Schematically,
\beq
\int
{\cal D}\Psi
{\cal D}\bar{\Psi}
\exp
\left[
i\int d^4x
\bar{\Psi}
\left(
i\not{\partial} - m_f - y\phi
\right)
\Psi
\right]
=
\det
\left(
i\not{\partial}
- m_f - y\phi
\right).
\eeq
Using the relation $\det A=\exp({\rm Tr}\log A)$, the fermionic determinant contributes to the scalar effective action. 
The low energy scalar EFT below the fermion threshold can therefore be written as
\beq
U_{\rm EFT}(\phi)
=
t_{\rm EFT}\phi
+
\frac{m_{\rm EFT}^2}{2}\phi^2
+
\frac{a_{\rm EFT}}{3}\phi^3
+
\frac{\lambda_{\rm EFT}}{4}\phi^4
+
\sum_{n>4}
c_n
\frac{\phi^n}{m_f^{n-4}},
\eeq
where the coefficients $c_n$ contain the appropriate powers of $y$. For the nonderivative background-field expansion considered here, higher-order terms are suppressed by powers of $y\phi/m_f$.

The EFT coefficients are determined by matching the low-energy Green functions
$\Gamma^{(n)}$
calculated in the full theory (i.e., scalar + fermion) and in the scalar-only EFT. At the matching scale, the condition can be written schematically as
\beq
\Gamma_{\rm EFT}^{(n)}
\left(
p_i\ll m_f
\right)
=
\Gamma_{\rm full}^{(n)}
\left(
p_i\ll m_f
\right),
\eeq
where $n=1,2,3,4$ correspond to the scalar one-point, two-point, three-point, and four-point functions, respectively.
In the present analysis, we restrict the matching to the nonderivative operators in the scalar effective potential, corresponding to the zero external momentum limit of the scalar Green functions.

For example, the scalar two-point function in the full theory contains a virtual fermion-loop contribution. The corresponding fermion loop is absent in the scalar-only EFT. Its low-energy effect must therefore be reproduced by the EFT mass parameter. This gives the matching relation
\beq
m_{\rm EFT}^2(m_f)
=
m_R^2(m_f)
+
\delta m_{\rm th}^2
\eeq
where $\delta m_{\rm th}^2$ is the finite heavy-fermion threshold correction~\cite{Braathen:2018htl}.

The coefficient of the linear term in the one-loop potential at $\mu=m_f$ gives the tadpole threshold correction
\beq
\delta t_{\rm th}
=
\frac{ym_f^3}{4\pi^2}.
\eeq

The coefficient of the quadratic term satisfies
\beq
\delta m_{\rm th}^2
=
\frac{y^2m_f^2}{4\pi^2},
\eeq
while the cubic term in the one-loop potential gives
\beq
\delta a_{\rm th}
=
-\frac{y^3m_f}{2\pi^2}.
\eeq
Similarly,
\beq
\delta\lambda_{\rm direct}
=
-\frac{2y^4}{3\pi^2}.
\eeq
After the ultraviolet-divergent parts of the fermion-loop contributions are cancelled by the corresponding counterterms in the $\overline{\rm MS}$ scheme, the remaining finite pieces define the heavy-fermion threshold corrections. Thus, the counterterm contributions are implicit in the finite quantities $\delta m_{\rm th}^2$, $\delta a_{\rm th}$, and $\delta\lambda_{\rm direct}$ appearing in the matching relations below.
The resulting threshold matching relations are
\beqa
m_{\rm EFT}^2(m_f)
=&&
m_R^2(m_f)
+
\frac{y^2m_f^2}{4\pi^2},~
a_{\rm EFT}(m_f)
= a_R(m_f)-
\frac{y^3m_f}{2\pi^2},
\text{and} \no
\lambda_{\rm EFT}(m_f)
=&& 
\lambda_R(m_f)-
\frac{2y^4}{3\pi^2}.
\eeqa

The tadpole requires a separate vacuum renormalization condition. Threshold matching determines how the parameters of the full theory and the EFT are related, whereas the vacuum renormalization condition determines the field value around which the renormalized theory is expanded.
We choose $\phi=0$ as the renormalized vacuum and impose
$
U_{\rm eff}'(0)
=0
$.
We also choose the renormalized expansion point such that $t_R(mf)=0$.
The finite part of the tadpole counterterm is chosen so that
$\delta t_{\rm CT}
+
\delta t_{\rm th}
= 0
$.
Therefore,
\beq
\delta t_{\rm CT}
=
-\frac{ym_f^3}{4\pi^2}
\eeq
and the linear coefficient in the low-energy potential satisfies
$
t_{\rm EFT}
= 0.
$

Below the matching scale, the heavy fermion is no longer an active degree of freedom in the EFT. The scalar EFT coefficients then evolve according to the scalar-theory renormalization-group equations,
\beq
\mu
\frac{d m_{\rm EFT}^2}{d\mu}
=
\beta_{m^2}^{\rm EFT},~
\mu
\frac{d a_{\rm EFT}}{d\mu}
=
\beta_a^{\rm EFT}
\eeq
and
\beq
\mu
\frac{d\lambda_{\rm EFT}}{d\mu}
=
\beta_\lambda^{\rm EFT}.
\eeq
The threshold matching conditions determine the boundary values of these coefficients at $\mu=m_f$, while the EFT renormalization-group equations determine their evolution toward lower scales.

In the parameter region relevant to the present SIULDM scenario, the quartic coupling $\lambda_{\rm EFT}$ is extremely small and the Yukawa coupling allowed by the quadratic naturalness condition derived below is also very small. The scalar-only RG evolution below $m_f$ is therefore expected to be subleading compared with the finite heavy-fermion threshold corrections. 
As a simple estimate of the RG effect below the matching scale, consider the one-loop running of the quartic coupling in the scalar-only EFT.  The one-loop beta function is~\cite{Machacek:1984zw}
\beq
\mu\frac{d\lambda_{\rm EFT}}{d\mu}
=
\frac{9}{8\pi^2}\lambda_{\rm EFT}^2.
\eeq
For the benchmark values \(m=10^{-22}\,{\rm eV}\) and \(\tilde m=5\,{\rm eV}\), the quartic coupling is
\beq
\lambda_{\rm EFT}
\simeq
\left(\frac{m}{\tilde m}\right)^4
\simeq
1.6\times10^{-91}.
\eeq
Running from the electron threshold \(m_f=0.511\,{\rm MeV}=5.11\times10^5\,{\rm eV}\) down to \(\tilde m=5\,{\rm eV}\) gives
$
\log\left(\frac{m_f}{\tilde m}\right)
\simeq
11.53.
$
To leading-logarithmic accuracy, the fractional change in the quartic coupling is therefore~\cite{Aghaie:2026bqq}
\beq
\left|
\frac{\delta\lambda_{\rm EFT}}{\lambda_{\rm EFT}}
\right|
\simeq
\frac{9\lambda_{\rm EFT}}{8\pi^2}
\log\left(\frac{m_f}{\tilde m}\right)
\simeq
2.1\times10^{-91}.
\eeq
As a more conservative check, we can run the coupling down to the infrared scale $\mu\sim m$. For $m=10^{-22}\rm{eV}$, one has $\ln(m_f/m)\simeq 63.80$, while the resulting fractional change remains only $\sim 1.16\times10^{-90}$.
Thus, despite the large hierarchy between the electron mass and the ULDM scale, the scalar-only RG running is completely negligible because the scalar self-coupling itself is extremely small. The running of the other scalar EFT coefficients is likewise highly suppressed in the radiatively natural parameter region. Therefore, using the threshold-matched coefficients without further RG evolution provides an excellent approximation for the present analysis.
We consequently use the threshold-matched coefficients as the leading approximation.

The loop-generated cubic interaction requires special consideration. Its leading contribution vanishes after first-order nonrelativistic time-averaging, but this does not imply that the cubic interaction has no effect on low-energy scalar scattering.
To see the first point, we write the real scalar field in the nonrelativistic form
\beq
\phi(t,\mathbf{x})
=
\frac{1}{\sqrt{2m}}
\left[
\Psi(t,\mathbf{x})e^{-imt}
+
\Psi^*(t,\mathbf{x})e^{imt}
\right]
\eeq
where $\Psi(t,\mathbf{x})$ changes slowly on the time scale $m^{-1}$.
The cubic power of the field is
\beqa
\phi^3
=
\frac{1}{(2m)^{3/2}}
\Big[
\Psi^3e^{-3imt}
+
3\Psi^2\Psi^*e^{-imt}
+
3\Psi(\Psi^*)^2e^{imt}
+
(\Psi^*)^3e^{3imt}
\Big]
\eeqa

Every term in this expression contains a rapidly oscillating phase. If the averaging interval $\Delta t$ satisfies
$
\Delta t
\gg
m^{-1}
$
then
$
\left\langle
e^{\pm imt}
\right\rangle
\simeq 0
$
and
$
\left\langle
e^{\pm3imt}
\right\rangle
\simeq 0.
$
Consequently,
\beq
\left\langle
\phi^3
\right\rangle
\simeq
0
\eeq
at first order in the nonrelativistic averaging.

However, two cubic vertices contribute at second order to scalar $2\rightarrow2$ scattering through an intermediate virtual scalar. Following Ref.~\cite{Arvanitaki:2014faa}, this effect can be represented as a correction to the effective nonrelativistic quartic coupling,
\beq
\lambda_{\rm NR}
= \lambda_{\rm EFT}
-
\frac{10a_{\rm EFT}^2}{9m_{\rm EFT}^2}.
\eeq
Since the scalar-only RG evolution is negligible, we identify the low-energy EFT coefficients with their threshold-matched values in the following.
We will use the approximation
$m_{\rm EFT}\simeq m$ from now on.

To isolate the cubic interaction generated by the Yukawa loop, we adopt the benchmark boundary condition
$
a_R(m_f)=0.
$
This condition is not enforced by a $\phi\rightarrow-\phi$ symmetry because the Yukawa interaction explicitly breaks that symmetry. It should therefore be understood as a boundary condition on the renormalized cubic coupling at the matching scale.
With this choice, the cubic coupling below the fermion threshold is
\beq
a_{\rm EFT}(m_f)
=
-\frac{y^3m_f}{2\pi^2}.
\eeq
The cubic-induced contribution to the effective quartic coupling is then
\beq
\delta\lambda_{\rm cubic}
=
-\frac{10a_{\rm EFT}^2(m_f)}{9m^2}
=
-\frac{5y^6m_f^2}{18\pi^4m^2}.
\eeq
The direct fermion-loop contribution to the quartic coupling is
\beq
\delta\lambda_{\rm direct}
=
-\frac{2y^4}{3\pi^2},
\eeq
which is obtained directly from the coefficient of $\phi^4$ in the fermionic one-loop effective potential. By contrast, $\delta\lambda_{\rm cubic}$ is generated indirectly by two insertions of the radiatively generated cubic interaction.
The effective low-energy quartic interaction is therefore
\beq
\lambda_{\rm NR}
=\lambda_{\rm EFT}+\delta \lambda_{\rm cubic}
=
\lambda_R
+
\delta\lambda_{\rm direct}
+
\delta\lambda_{\rm cubic}
= \lambda_R
- \frac{2y^4}{3\pi^2}
-
\frac{5y^6m_f^2}{18\pi^4 m^2}.
\eeq

We now reinterpret the radiative corrections in terms of radiative naturalness; the physical ULDM mass can remain small even when the threshold correction is larger than $m^2$, provided that the renormalized mass parameter is adjusted to cancel the correction. Such a cancellation is mathematically allowed, but an increasingly large correction requires increasingly precise tuning.
We define the quadratic naturalness measure as
\beq
\Delta_m
\equiv
\frac{
|\delta m_{\rm th}^2|
}{
m^2
}
=
\frac{
y^2m_f^2
}{
4\pi^2m^2
}.
\eeq

The reference condition $\Delta_m=1$ means that the heavy-fermion threshold correction is equal in magnitude to the physical ULDM mass-squared scale. This condition gives
the quadratic naturalness scale
\beq
\label{y1}
y_{\rm nat}^{(m)}
\equiv 
\frac{2\pi m}{m_f}.
\eeq
Values satisfying $y>y_{\rm nat}^{(m)}$ are not theoretically forbidden. Instead, they require an increasing cancellation between the renormalized scalar mass parameter $m_R^2$ and the threshold correction $\delta m_{\rm th}^2$.

We similarly define the quartic naturalness measure as
\beq
\Delta_\lambda
\equiv
\frac{
\left|
\delta\lambda_{\rm direct}
+
\delta\lambda_{\rm cubic}
\right|
}{\lambda}
=
\left|\frac{
2y^4
}{
3\pi^2\lambda}
+
\frac{
5y^6m_f^2
}{18\pi^4m^2\lambda}
\right|.
\eeq

At $y=y_{\rm nat}^{(m)}$, we compare the radiative quartic corrections with the low-energy quartic coupling $\lambda=(m/\tilde{m})^4$ required by the cosmological analysis in Section II.  Since
$
\lambda_{\rm NR}=
\lambda_R
+
\delta\lambda_{\rm direct}
+
\delta\lambda_{\rm cubic},
$
the cosmological requirement corresponds to $\lambda_{\rm NR}\simeq\lambda$.
The ratio of the direct quartic threshold correction to the cosmologically required quartic coupling is
\beq
\frac{
|\delta\lambda_{\rm direct}|
}{\lambda}
=
\frac{2y^4}{3\pi^2\lambda}
=
\frac{32\pi^2}{3}
\left(\frac{\tilde m}{m_f}
\right)^4
\simeq
9.7\times10^{-19}.
\eeq
The corresponding cubic-induced correction satisfies
\beq
\frac{|\delta\lambda_{\rm cubic}|
}{\lambda}
=
\frac{5y^6m_f^2}{ 18\pi^4m^2\lambda}
=
\frac{160\pi^2}{9}
\left(\frac{\tilde m}{m_f}
\right)^4
\simeq
1.6\times10^{-18}.
\eeq

Thus, throughout the parameter region satisfying the quadratic naturalness criterion
$(y\le y_{\rm nat}^{(m)})$,
both radiative corrections are much smaller than the cosmologically required quartic coupling. It therefore follows that
\beq
\lambda_R
\simeq
\lambda_{\rm EFT}
\simeq
\lambda_{\rm NR}
\simeq \lambda
\eeq
to very high accuracy.
Therefore, the cosmological constraints derived in Section II in terms of the quartic coupling \(\lambda\) remain essentially unchanged in the radiatively natural parameter region
($\Delta_m \le 1$).

It is also necessary to verify that the loop-generated cubic interaction does not significantly modify the homogeneous cosmological evolution before the nonrelativistic regime is reached.
For the boundary condition $a_R(m_f)=0$, the loop-generated cubic coefficient 
at $y=y_{\rm nat}^{(m)}$ becomes
\beq
a_{\rm EFT}
=
-\frac{
4\pi m^3}{
m_f^2}.
\label{aeft2}
\eeq
The ratio of the cubic contribution to the quadratic contribution at $\phi=\phi_m$ is defined as
\beq
R_{32}(\phi_m)
\equiv
\frac{
|a_{\rm EFT}\phi_m^3|/3
}{
m^2\phi_m^2/2
}
=\frac{8\sqrt{2}\pi}{3}\left(\frac{\tilde m}{m_f}\right)^2,
\eeq
where we have
used Eq. (\ref{phim})
and Eq. (\ref{aeft2}).
For $\tilde m=5~{\rm eV}$ and $m_f=5.11\times10^5~{\rm eV}$, we obtain
\beq
R_{32}(\phi_m)
\simeq
1.1\times10^{-9}\ll 1,
\eeq
which implies
that we can neglect the
cubic term compared to
the quadratic term
for 
$y\le y_{\rm nat}^{(m)}$,
since $|a_{\rm EFT}|\sim y^3$.

At $\phi=\phi_m$, the quadratic and quartic contributions have equal magnitudes by definition. Therefore, the ratio of the cubic contribution to the quartic contribution at the transition point 
$R_{34}(\phi_m)$ is equal to $R_{32}(\phi_m)$.
For a  general
amplitude $|\phi|$,
\beq
R_{34}(\phi)
\equiv
\frac{
|a_{\rm EFT}\phi^3|/3
}{
\lambda\phi^4/4
}=\frac{8\sqrt{2}\pi}{3}\left(\frac{\tilde m}{m_f}\right)^2\frac{\phi_m}{\lvert\phi\rvert}
=
R_{32}(\phi_m)\frac{\phi_m}{\lvert\phi\rvert}
\label{R34}
\eeq
at $y=y^{(m)}_{\rm nat}$.
This ratio decreases as $|\phi|$ increases. Hence the loop-generated cubic interaction does not significantly modify the quartic-dominated cosmological evolution in the parameter region satisfying the quadratic naturalness condition.

Finally, we need to check that the cubic term is also negligible compared to the quartic term on galactic scales. In Phase III, the ULDM energy density near the Solar System is $m^2 \phi^2/2\sim {\rm GeV}/{\rm cm}^3\sim 10^{-42}{\rm GeV}^4$. This leads to a local field amplitude  $|\phi|\sim   10^{-21}{\rm GeV}^2/m$, and hence
from Eq. (\ref{R34})
and Eq. (\ref{phim}),
\beq
\label{R34phi}
R_{34}(\phi)
=
\frac{16\pi}{3}
\frac{\tilde m^4}
{m_f^2 m\lvert\phi\rvert}
\sim 10^{4}
\left(\frac{\tilde m}{m_f}\right)^2
\left(\frac{\tilde m}{\rm eV}\right)^2
\ll 1.
\eeq
The value of $R_{34}(\phi)$ in Eq.~(\ref{R34phi}) is maximal within the
radiatively natural region at $y=y_{\rm nat}^{(m)}$; for smaller
values of $y$, it decreases as $y^3$.
Therefore, we can also safely apply the constraints from galactic dynamics discussed in Section II
when the quadratic naturalness criterion
is satisfied.

The radiative corrections obtained above should therefore not be interpreted as absolute theoretical exclusion bounds. Instead, they quantify the degree of cancellation required to maintain the extremely small ULDM parameters in the presence of a coupling to ordinary matter. In the parameter region of interest, the scalar mass threshold correction provides the strongest radiative naturalness criterion. Once the quadratic naturalness condition is satisfied, both the direct quartic correction and the cubic-induced quartic correction are negligibly small compared with the repulsive self-interaction required by the SIULDM cosmology.
For $\tilde{m}=5~{\rm eV}$, using Eq. (\ref{phiosc1}), we obtain
$
\phi_{\rm osc} \simeq 6.7 \times 10^{26}~{\rm eV}.
$
The condition 
$
\frac{y\phi_{\rm osc}}{m_f} \lesssim 1
$
corresponds approximately to
$
y \lesssim 7.6 \times 10^{-22}.
$

\begin{figure*}[tb]
\centering
\includegraphics[width=0.49\textwidth]{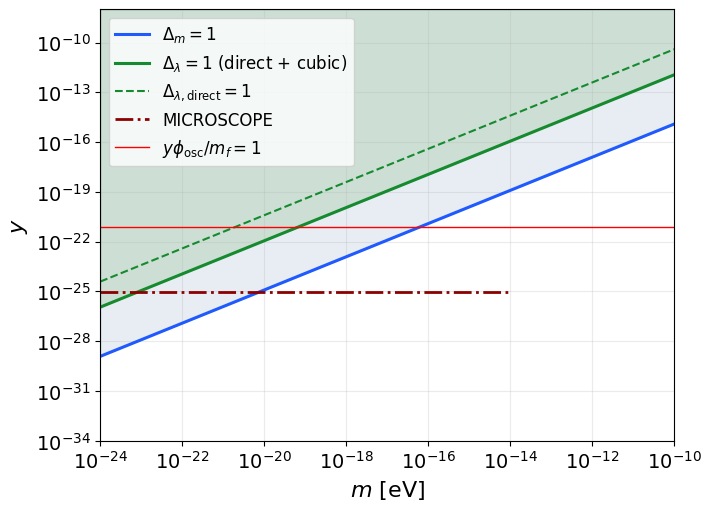}
\caption{
Radiative naturalness constraints on the Yukawa coupling $y$. The heavy fermion is matched onto the low-energy scalar EFT at the threshold scale $\mu=m_f$. We use $m_f=0.511~{\rm MeV}$ and $\tilde m=5~{\rm eV}$. The curves represent contours of the naturalness measures $\Delta_m$ and $\Delta_\lambda$. The shaded regions with $\Delta_m>1$ or $\Delta_\lambda>1$ are not theoretically excluded. Instead, they correspond to increasing cancellation between the renormalized parameters and the radiative threshold corrections. 
The direct+cubic quartic contour assumes the benchmark boundary condition \(a_R(m_f)=0\).
The red dashed line labeled MICROSCOPE represents the upper bound on the Yukawa coupling  obtained by translating the MICROSCOPE constraint on the electron-mass coupling $d_{m_e}$
~\cite{Baryakhtar:2025uxs}.
The red solid line marks the indicative boundary \(y\phi_{\rm osc}/m_f=1\). The region above it lies outside the regime of validity of the truncated background-field expansion.
}
\label{d}
\end{figure*}

One possible way to probe ULDM is through precision frequency measurements using atomic clocks~\cite{Kouvaris:2019nzd,Filzinger:2023zrs} or nuclear clocks~\cite{decol2026thorium229opticalnuclearclock}. The dilatonic coupling to a fermion mass is conventionally written as~\cite{Arvanitaki:2014faa}
\beq
{\cal L}_{\rm int}
=
-\kappa m_f d_{m_f}
\phi\bar{\Psi}\Psi
\eeq
where
$
\kappa
=
\frac{
\sqrt{4\pi}
}{M_{\rm Pl}}
$.

Comparing the dilatonic interaction with the Yukawa interaction
$
{\cal L}_Y
=
-y\phi\bar{\Psi}\Psi
$
gives
\beq
y
=
\kappa m_f d_{m_f}.
\eeq

For $m_f=m_e$, the interaction considered in the present model therefore corresponds specifically to the electron-mass coupling $d_{m_e}$ as
\beq
y=1.4848\times10^{-22}d_{m_e}.
\eeq
The MICROSCOPE experiment provides a stringent constraint on the linear coupling of an ultralight scalar field to the electron mass. The final MICROSCOPE result implies
~\cite{Baryakhtar:2025uxs}
\beq
d_{m_e} \lesssim 6\times10^{-4}.
\eeq
We translate this limit into a bound on $y$  in Fig.~2.
The MICROSCOPE bound lies above the quadratic naturalness curve for sufficiently small ULDM masses, indicating that the experimentally allowed Yukawa coupling can still require substantial radiative tuning of the ultralight scalar mass. In particular, for the benchmark mass \(m\sim10^{-22}\,{\rm eV}\), the quadratic naturalness condition is considerably stronger than the current MICROSCOPE constraint. Thus, MICROSCOPE does not yet probe the radiatively natural region associated with the scalar-mass correction at this mass scale.

As \(m\) increases, the quadratic naturalness bound scales approximately linearly with \(m\) and eventually intersects the nearly mass-independent MICROSCOPE limit
in the long-range regime $m r_{\oplus} \ll 1$ ~\cite{Baryakhtar:2025uxs}, where
$r_{\oplus}$ is the radius of the Earth.
Above this intersection, the experimental constraint becomes stronger than the quadratic naturalness criterion. 
The quartic naturalness bounds remain substantially weaker than the quadratic naturalness bound throughout the parameter range shown.

Experimental sensitivity curves used for comparison must consequently correspond to observables that are sensitive to $d_{m_e}$. Optical-optical clock comparisons that predominantly probe the photon coupling $d_e$ cannot be directly identified with the Yukawa interaction considered here. A consistent comparison instead requires sensitivity coefficients appropriate to the electron-mass coupling, such as those obtained from microwave-optical, atom-cavity, or related frequency comparisons.

The main implication of the present analysis is therefore not that Yukawa couplings above a particular value are forbidden by a renormalization-scale-independent stability condition. Instead, integrating out the heavy Standard Model fermion generates finite threshold corrections to the SIULDM mass and self-interactions. Because the physical ULDM mass is extremely small, the quadratic threshold correction can become much larger than $m^2$ even for a very weak Yukawa coupling. Values satisfying $y>y_{\rm nat}^{(m)}$ therefore require increasing radiative tuning unless an additional symmetry or ultraviolet mechanism protects the ultralight scalar mass. At the same time, throughout the radiatively natural parameter region, both the direct quartic correction and the cubic-induced effective quartic interaction remain much smaller than the repulsive quartic interaction required by the SIULDM cosmology.

\section{Discussion}
In this work, we have investigated the radiative naturalness of self-interacting ultralight dark matter coupled to Standard Model fermions. Cosmological evolution and galactic phenomenology favor an effective scale $\tilde m=m/\lambda^{1/4}$ of order a few {\rm eV}. In this regime, the quartic interaction controls the early oscillatory phase, while the quadratic mass term later dominates and allows the field to behave as CDM.

Integrating out a heavy fermion coupled through a linear Yukawa interaction generates finite threshold corrections to the scalar mass and self-interaction coefficients. We formulate these effects in a low-energy scalar EFT matched near $\mu\simeq m_f$ and interpret the resulting conditions as radiative naturalness criteria. The quadratic mass correction provides the strongest criterion because the physical ULDM mass is extremely small. By contrast, throughout the region satisfying the quadratic naturalness condition, both the direct quartic correction and the cubic-induced quartic contribution remain much smaller than 
the repulsive quartic interaction required by the SIULDM cosmology. The scalar-only RG running below the matching scale is also negligible.

The loop-generated odd-power interactions are treated consistently by imposing a vacuum renormalization condition for the tadpole and retaining the cubic contribution to the nonrelativistic quartic interaction. Their effect remains negligible in the radiatively natural region and does not significantly modify the cosmological or galactic dynamics considered here.
Our conclusion that the cubic-induced quartic correction remains subdominant
compared to the tree-level quartic term assumes the benchmark boundary condition $a_R(m_f)=0$ at the matching scale.

For the electron case, the Yukawa coupling can be mapped onto the electron-mass coupling $d_{m_e}$, allowing direct comparison with the MICROSCOPE constraint. Around $m\sim10^{-22}{\rm eV}$, the quadratic naturalness criterion is substantially stronger than the current  MICROSCOPE  bound, while at larger masses the experimental constraint can become more restrictive. The quartic naturalness bounds remain much weaker over the parameter range considered.

Our results therefore show that precision searches for ULDM should be interpreted together with the radiative structure of the scalar theory. Larger Yukawa couplings are not strictly excluded, but they require increasing cancellation among renormalized parameters and threshold corrections unless an additional ultraviolet mechanism protects the ultralight scalar mass. Future work should refine the cosmological treatment, the real-scalar galactic dynamics, and laboratory sensitivities to fermion-mass couplings.

\section*{Acknowledgments}
This work was supported in part by the U.S. Department of Energy (Grant No. DE-FG02-03ER41260). This research
used resources of the National Energy Research Scientific Computing Center, a DOE Office of Science User Facility
supported by the Office of Science of the U.S. Department of Energy under Contract No. DE-AC02-05CH11231 using
NERSC award NP-ERCAP0027381. CRJ thanks for the hospitality during his visit to the Asia Pacific Center for
Theoretical Physics (APCTP) where this work was completed. 


\end{document}